# Agent-based modelling of pedestrian responses during flood emergency: mobility behavioural rules and implications for flood risk analysis

Mohammad Shirvani[1], Georges Kesserwani[1*], Paul Richmond[2]

[1]*Department of Civil and Structural Engineering, University of Sheffield, Mappin St, Sheffield City Centre, Sheffield S1 3JD, UK*
[2]*Department of Computer Science, University of Sheffield, Mappin St, Sheffield City Centre, Sheffield S1 3JD, UK*

**Abstract**

An agent-based model (ABM) for simulating the interactions between flooding and pedestrians is augmented to more realistically model responses of evacuees during floodwater flow. In this version of the ABM, the crowd of pedestrians has different body height and weight, and extra behavioural rules are added to incorporate pedestrians' states of stability and walking speed in floodwater. The augmented ABM is applied to replicate an evacuation scenario for a synthetic test case of a flooded shopping centre. Simulation runs are performed with increasingly sophisticated configuration modes for the pedestrians' behavioural rules. Simulation results are analysed based on spatial and temporal indicators informing on the dynamic variations of the flood risk states of the flooded pedestrians, i.e. in terms of a commonly used flood Hazard Rating (HR) metric, variable walking speed, and instability due to toppling and/or sliding. Our analysis reveal significantly prolonged evacuation times and risk exposure levels as the stability and walking speed behavioural rules become more sophisticated. Also, it allows to identify more conservative HR thresholds of pedestrian instability in floodwater, and a new formula relating walking speed states to HR for stable pedestrian in floodwater. Accompanying details for software accessibility are provided.

**Short title:** Modelling people mobility in floodwater

*Corresponding author.*
*Georges Kesserwani, Department of Civil and Structural Engineering, University of Sheffield, Mappin St, Sheffield City Centre, Sheffield S1 3JD, UK*
*Email: g.kesserwani@sheffield.ac.uk*

# 1. Introduction

Evacuation simulation models are useful tools to support analysis of flood risk to people under immediate emergency condition in small urban areas, i.e. less than 0.5 km × 0.5 km (e.g. Dawson et al. 2011; Lumbroso & Davison 2018). These models serve various purposes, such as emergency response management (Lumbroso & Tagg 2011), evacuation route finding (Pillac et al. 2016; Bernardini et al. 2017b), and assessing the number of casualties (Lumbroso & Davison 2018). The motion of individuals, their interactions with each other and with their surrounding environment, such as buildings, obstacles and walls, are the key elements in simulating crowds' evacuation processes (Bernardini et al. 2016; Bernardini et al. 2017a; Makinoshima et al. 2018; Alvarez & Alonso 2018; Lumbroso & Davison 2018). Most of these models use a local motion planning model (i.e. *social force model*) for representation of people's evacuation, which is commonly integrated within agent-based modelling (ABMs) platforms. This is mainly because of the capability of ABMs in bridging between physical dynamics of evacuation processes and involving mitigation policies that makes them suitable tools for food resilience studies (Lumbroso & Davison 2018). However, most of these ABMs are particularly designed for pre-evacuation planning prior to flooding, where the main focus is given to static evacuation parameters (Bernardini et al. 2017b; Alvarez & Alonso 2018; Bernardini et al. 2020). More specifically, these models overlook potential changes in local evacuees' responses to transient changes in the hydrodynamics of floodwater (Bernardini et al. 2017b; Alvarez & Alonso 2018; Bernardini et al. 2020).

To analyse flood risk on people under immediate evacuation condition in the presence of flowing floodwater, few models considered dynamic coupling with a hydrodynamic model. One of these models is the Life Safety Model (LSM, www.lifesafetymodel.net) that enables estimation of evacuation time and the number of injuries and casualties based on local information of floodwater on each person (Lumbroso & Di Mauro 2008). In the LSM model, the instability and drowning state of people are used as a metric to quantify the flood risk on people and to control their mobility in floodwater (Lumbroso & Davison 2018). The stability threshold for people is defined by combination of floodwater depth $d$ (m) and depth-averaged velocity magnitude $v$ (m/s) for a person of average height and weight according to experimental data (0.6 m²/s ≤ $d \times v$ ≤ 1.7 m²/s). Also, the LSM is limited to considering uniform height and weight for all the people alongside a constant walking speed in floodwater. Another available model is called FloodPEDS developed by Bernardini *et al.* (2017a). This model uses similar stability state of evacuees in floodwater as a metric to quantify the flood risk to people. The mobility of people in FloodPEDS is also controlled by their stability state, i.e. defined based on whether they are exposed to extreme floodwaters with $d \times v \geq 1.2$ m²/s or not. This model benefits from an experimental-based modified *social force model* (SFM) to include more realistic behavioural aspects for evacuees such as their variable motion speeds in floodwater and path choices. Both LSM and FloodPEDS are not designed to take into account people's different body height and weight. More recently, a flood-pedestrian simulator was developed by Shirvani *et al.* (2020) that dynamically couples a hydrodynamic ABM to a pedestrian ABM. This simulator captures both-way interactions across the flooding and pedestrian dynamics in space and time, and has been evaluated for emergency evacuation and a community intervention to deploy a temporary flood barrier (Shirvani *et al.* 2020). This version of the simulator is still limited to using constant walking speed thresholds for pedestrians in floodwater that were assumed related to the flood HR thresholds (see Table 1) published in official UK guidance (Environment Agency 2006), and excludes any behavioural rules to represent unstable pedestrians. An augmented version is therefore desired that concurrently incorporates variable walking speed and stability conditions behavioural rules for the pedestrians in floodwater, while considering them with diverse body height and mass.

In recent years, experimental research efforts have been made to determine empirical formulae for approximating people stability (e.g. Russo *et al.* 2013; Xia *et al.* 2014) and walking speed (e.g. Ishigaki *et al.* 2009; Postacchini *et al.* 2018; Lee *et al.* 2019; Bernardini *et al.* 2020) in floodwater at laboratory-scale. Russo *et al.* (2013) conducted experiments on human subjects to identify thresholds of people stability in relation to water depth and velocity. Xia *et al.* (2014) used partially submerged subjects in laboratory flumes to identify formulae for estimation of incipient velocities at the threshold of toppling and sliding states of individuals based on corresponding water depth and human body characteristics. Ishigaki et al. (2009) used laboratory flumes to relate human walking speed to an estimated specific force based on water depth and velocity with application to underground flooding for safe evacuation planning. Postacchini et al. (2018) conducted a set of experiments to estimate the speed of people in relation to the specific force following the work of Ishigaki et al. (2009). They provided an empirical formula for estimating variable people walking speed in floodwater. Later on, Lee et al. (2019) quantified the walking and running speed of assisted and unassisted elderly people in swimming pools with water depths of 10, 20, 30, 40 and 50 cm. Their study identifies a significant difference in the evacuation speed of the elderly relative to that of younger ones moving in the same depth of water. More recently, Bernardini et al. (2020) further improved the formulae of Postacchini et al. (2018) to further account for people's gender, age and body characteristics. However, such formulae are yet to be implemented as behavioural rules in pedestrian evacuation models to assess the extent to which they can improve evacuation planning strategies in flooding emergencies.

This paper explores the relevance of increasing the level of sophistication of the behavioural rules governing the mobility of individual pedestrians in floodwater for flood risk analysis. This exploration is facilitated by augmenting the functionality of a flood-pedestrian simulator that couples validated hydrodynamic and pedestrian ABMs (Section 2.1). The simulator is augmented so that it concurrently outputs spatial and temporal outcomes on: the risk states of pedestrians in floodwater in relation to the HR metric (Section 2.1), and their states of walking speed when stable or otherwise their unstable states due to toppling and sliding (Section 2.2). The augmented simulator incorporates randomised pedestrians with various body height and weight, and experimentally valid behavioural rules across the ABMs in integrating walking speed states or unstable states of individual pedestrians in floodwater. The augmented simulator is applied to replicate synthetic test case of an emergency evacuation of one thousand pedestrians during a worst-case scenario flooding (Section 2.3). The simulator runs are applied diagnostically under three configuration modes, while increasing the level of sophistication of the behaviour rules, to systematically evaluate the relative changes in the outcomes across the modes (Section 3.1). The outcomes of simulator are discussed considering practical implications for flood risk assessment based on focused analysis to the pedestrian response dynamics, and outlaying limitations (Section 3.2). Conclusions are drawn (Section 4) reflecting on the future research and details of software accessibility are provided in the acknowledgement section.

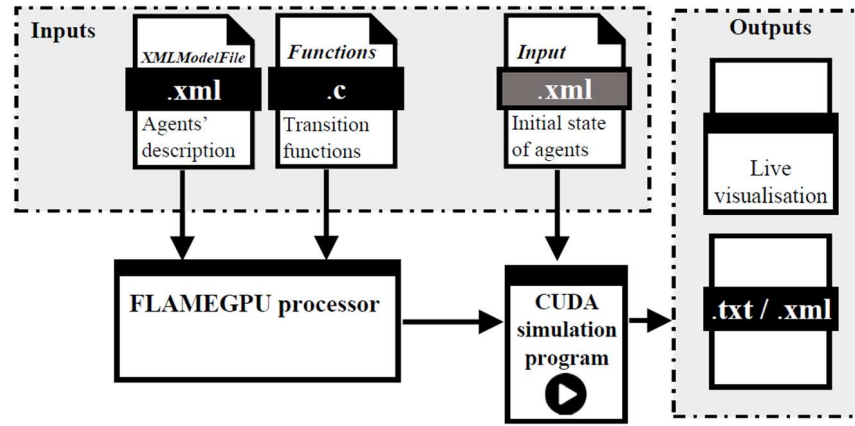

**Figure 1.** The process for generating and running an agent-based simulation program on FLAMEGPU (http://www.flamegpu.com/home). A detailed list of the agents' description and initial states is available in the accompanying 'run guide' document of the flood-pedestrian simulator software (see also the Acknowledgements section).

## 2. Materials and methods

### 2.1 Flood-pedestrian simulator

The flood-pedestrian simulator couples a hydrodynamic ABM to a pedestrian ABM for modelling the two-way interactions between flooding and individuals (Shirvani *et al.* 2020). It is implemented via the Flexible Large-scale Agent-based Modelling Environment for the GPU (FLAMEGPU), which allows the simulation of multiple agent interactions on CUDA cores of a Graphical Processing Unit (GPU) (Richmond *et al.* 2009; Chimeh & Richmond 2018). FLAMEGPU platform allows to create and run a CUDA simulation program by processing three inputs (Figure 1). The *XMLModelFile.xml* is where a user defines formal agent specifications, including their descriptive information, type, numbers, properties, *etc*. An agent can be specified in space as either *discrete* or *continuous* (FLAMEGPU Documentation and User Guide). *Discrete agents* have fixed coordinates and must be pre-allocated in the memory of the GPU as 2D grid of size of a power of two number (e.g. 64 × 64, 128 × 128, 256 × 256, 512 × 512, *etc.*). *Continuous agents* change their coordinates and their population; they can be of any number as much as the GPU memory can accommodate. The *input.xml* file contains the initial conditions of the variables of state of all the defined agents. In a *single C script*, the behaviour rules to update all agents are implemented, and includes *Transition functions* to achieve dynamic message communication across multiple agents as they get simultaneously updated (FLAMEGPU Documentation and User Guide).

The pedestrian ABM on FLAMEGPU (Karmakharm *et al.* 2010), adopts a validated SFM used for modelling pedestrian flow and evacuation dynamics on dry study domains (Helbing & Molnár 1995; Helbing *et al.* 2000). Within the pedestrian ABM, there are two specified types of agent: *navigation agent* and *pedestrian agent*. Navigation agents are defined to be discrete on a grid size of 128 × 128 spanning a *navigation map* on the study domain. Each navigation agent stores the information required for steering and directing one or more moving pedestrian(s) located over it towards their goal destination, such as the location of the entrances, exits, terrain obstacles and walls. Pedestrian agents are of continuous type which allows them to change coordinates in space and over time. Pedestrian agents receive the steering and directional information as a message from the navigation agent as they move over the navigation grid.

In the hydrodynamic ABM, there is only one type of agents, discrete *flood agents*. They form a two-dimensional grid that is aligned with the grid of navigation agents. Each flood agent has a fixed coordinate on the grid and it stores terrain properties in terms of height $z$ (m) and roughness parameter $n_M$ (s m$^{-1/3}$), and states of floodwater variables in terms of water depth $d$ (m) and velocity components $v_x$ (m/s) and $v_y$ (m/s). The states of floodwater variables stored in all the flood agents are simultaneously updated via a *non-sequential* implementation (Shirvani *et al.* 2020) of a robust finite volume numerical solver for the shallow water equations that is validated for real-world applications (Wang *et al.* 2011).

The coupling between the pedestrian and hydrodynamic ABMs is achieved via using the grid of navigation agents as an intermediate, where ad-hoc messages are allocated to passed information between the two models. Each navigation agent is set to receive the floodwater variables (i.e. $d$, $v_x$ and $v_y$) from the flood agent at its aligned location. The recipient navigation agent subsequently processes $d$ and a velocity magnitude $v = \max(|v_x|, |v_y|)$ into a flood Hazard Rating (HR), where HR = $(v + 0.5) \times d$ (Kvočka et al. 2016; Willis et al. 2019; Costabile et al. 2020) in line with the risk-to-people method developed for the UK Environment Agency (2006). Once HR is estimated by the navigation agent, it is sent to any pedestrian agent within its area via a message (Shirvani et al. 2020). Responding to these messages, all the pedestrian agents are set to act as evacuees once a positive HR value is received by any of their counterpart. The evacuating behaviour of pedestrian agents follows two rules: they will no longer enter the domain, and those remaining in the domain will move immediately to a pre-defined emergency exit regardless of their pre-planned destination. In wet areas, a set of rules governing the movement of pedestrians in floodwater is implemented based on four *walking speed states* according to HR-related *flood-risk state* as described in Table 1 (Shirvani et al. 2020). The motion speed of pedestrian agents in dry area is set to maintain 1.4 m/s, representative of an average human walking speed (Wirtz & Ries 1992; Mohler et al. 2007), but this speed may slightly increase or decrease for collision avoidance with each other or with obstacles.

The flood-pedestrian simulator is further augmented to characterise pedestrian agents with randomised different body height and mass; and, to realistically represent their variable motion speed and mobility states based on new experimentally-valid behavioural rules (Section 2.2).

**Table 1.** *Walking speed* and *flood risk states* of individual pedestrians *in floodwater* according to the ranges HR reported in the risk to people method (Environment Agency 2006; Kvočka et al. 2016).

| HR range | | Flood risk state | Walking speed state |
|---|---|---|---|
| 0 | 0.75 | 'Low' (safe for all) | 1.8 m/s (brisk walk to rush evacuation) |
| 0.75 | 1.5 | 'Medium' (danger for some, i.e. children) | 1.8 m/s (brisk walk to rush evacuation) |
| 1.5 | 2.5 | 'High' (danger for most) | 1.0 m/s (slow walk hindered by floodwater) |
| 2.5 | 20 | 'Severe' (danger for all) | 0.0 m/s (cannot walk due to severe floodwater) |

**2.2 Augmented version**

**Different body height and mass.** To take into account variations of people's body characteristics in the simulations, the pedestrian ABM is characterised with new functionality to generate pedestrian agents with different body height and mass. Each pedestrian agent is now randomly given a body height within the ranges shown in Figure 2. These ranges are based on the distribution of body height documented in the world data report (Roser et al. 2019). For this study, pedestrians shorter than 140 cm are excluded, assuming that they are kids that would be carried by adults. Pedestrians in the range of 140-163 cm are children who could not be carried, with a body mass estimated to $m_p = (l_p)^2 \times BMI$ (Disabled World 2019) where $BMI$ (kg / m$^2$) denotes the body mass index (here taken 21.7 as an ideal average for children) and $l_p$ (m) is the body height. Any pedestrian agent above 163 cm is considered as an adult and their body mass ($m_p$) is estimated by the formula of Kokong et al. (2018), which is: $m_p = [(0.01 \times l_p) - 1] \times 100$. This formula has been used with a 10% randomised uncertainty to account for deviations in the estimated $m_p$.

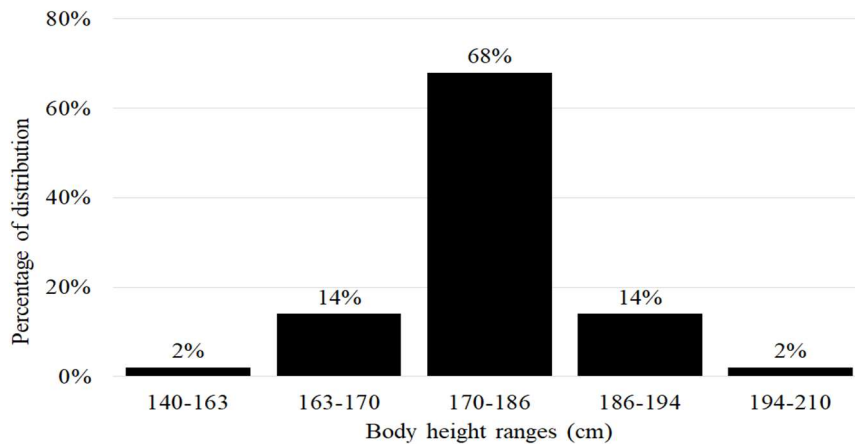

**Figure 2.** Body height distribution of pedestrians characterised according to the world's data report.

**Variable walking speed state.** To take into account realistic motion of individuals in floodwater, the empirical formula proposed by Bernardini et al. (2017b) and Postacchini et al. (2018) has been incorporated within the pedestrian ABM, as new behavioural rules to govern motion speed of pedestrian agents. These rules are only effective when pedestrian agents encounter non-zero water depth at their location. The incorporated formula is experimentally determined via estimating the minimum evacuation speed of a number of real subjects asked to walk as fast as possible in a laboratory flume (Postacchini et al. 2018). The physical characteristics of the participating individuals closely corresponds to the mean body height (175.3 cm for men and 161.9 cm for women) and weight (84.0 kg for men and 69.0 kg for women) of the UK population (Moody 2012). Denoting $V_i$ to be the walking speed of each pedestrian agent $i$, the empirical formula reads:

$$V_i = 0.53 M^{-0.19} \qquad (1)$$

$M$ is a function of specific force per width unit calculated based on the water depth $d$ and the velocity magnitude $v$, $M = \frac{v^2 d}{g} + \frac{d^2}{2}$, with $g$ is the gravitational constant. Like with HR (see in Section 2.1), each pedestrian agent $i$ processes the flood information that it receives from the navigation agent at its location to evaluate $M$, and use it to estimate and adopt a motion speed $V_i$ via Eq. (1).

**Stability states due to toppling and sliding.** The two experimentally-derived formulae for by Xia *et al.* (2014) are incorporated in the pedestrian ABM. These formulae are computed for each pedestrian agent independently to find the incipient velocity limit, $U_c$, relevant to their stability state based on their specific body characteristics and floodwater depth at their location. These formulae combine multiple forces that lead to toppling and sliding of flooded individuals including buoyancy, drag, effective weight and frictional forces. The limits $U_c^{(toppling)}$ and $U_c^{(sliding)}$ beyond which a human body loses stability in floodwater are:

$$U_c^{(toppling)} = a \left(\frac{d}{h_p}\right)^{\beta} \sqrt{\frac{m_p}{\rho_f d^2} - \left(\frac{a_1}{h_p^2} - \frac{b_1}{d\, h_p}\right)(a_2 m_p + b_2)} \qquad (2)$$

$$U_c^{(sliding)} = a \left(\frac{d}{h_p}\right)^{\beta} \sqrt{\frac{m_p}{\rho_f h\, h_p} - \left(a_1 \frac{d}{h_p} + b_1\right)\frac{(a_2 m_p + b_2)}{h_p^2}} \qquad (3)$$

where $\rho_f$ (= 997 kg m$^{-3}$) is the density of water, $h_p$ (m) and $m_p$ (kg) stand for the height and mass of a human body, respectively, with $a_1 = 0.633$, $b_1 = 0.367$, $a_2 = 0.001015$, and $b_2 = 0.0004927$ being non-dimensional coefficients defining the characteristic parameters of the human body structure. The parameters $a = 3.472$ and $\beta = 0.188$ are related to human body shape, which were calibrated using laboratory experiments (Xia *et al.* 2014). Like HR and $M$, each pedestrian agent processes the flood information that it receives from the navigation agent at its location to evaluate $U_c^{(toppling)}$ and $U_c^{(sliding)}$, and then it adopts a stability state according to the conditions described in Table 2.

**Table 2.** The stability states of pedestrian agents in the flood-pedestrian simulator identified by comparing the toppling and sliding incipient velocities to the velocity magnitude of floodwater.

| Condition | Stability state of a pedestrian in floodwater |
|---|---|
| $v < U_c^{(toppling)}$ and $v < U_c^{(sliding)}$ | 'Stable' (with a variable walking speed) |
| $v > U_c^{(toppling)}$ and $v < U_c^{(sliding)}$ | 'Toppling' (zero walking speed) |
| $v < U_c^{(toppling)}$ and $v > U_c^{(sliding)}$ | 'Sliding' (zero walking speed) |
| $v > U_c^{(toppling)}$ and $v < U_c^{(sliding)}$ | 'Toppling and sliding' (zero walking speed) |

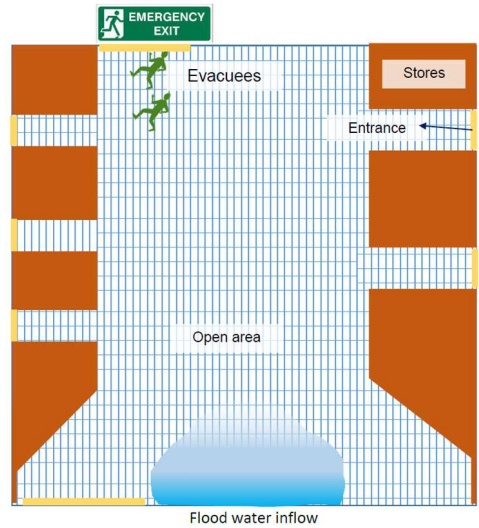

**Figure 3.** Sketch of the hypothetical shopping centre: the meshed area is where pedestrians can walk to the entrance doors that are coloured in 'yellow'. After the flood starts, all pedestrians will evacuate towards the 'emergency exit'. The blocks in 'brown' indicate terrain features and the 'blue' area in the southern part shows the location from where the flooding inflow breaches.

## 2.3 Synthetic test case and simulation set up

To explore the benefit of increasing level of sophistication in the behavioural rules governing the motion of pedestrian agents in floodwater, the synthetic test case of a flooded shopping centre under immediate evacuation scenario in Shirvani et al. (2020) is reconsidered. The test case involves a crowd of walking pedestrians that get exposed to a flash flood with no early warning nor early evacuation plan. It studies the behaviour of pedestrians before and during (as evacuees) the propagation of the floodwater while moving towards an emergency exit (Figure 3). The area of the shopping centre is 332 × 332 m², chosen considering the average area size of UK's 43 largest shopping centres (Gibson et al. 2018; Globaldata Consulting 2018; Sen Nag 2018; Tugba 2018). It includes stores, located at the east and west side, separated by corridors linking the entrance doors to an open area (Figure 3). Through these corridors, pedestrians can enter the open area and walk toward their destinations. The open area is assumed to be occupied by pedestrians, who can enter and leave to and from 7 entrance doors with an equal probability of one in seven. The flood propagation occurs from the southern side assuming a flood inundation condition. As flooding starts, pedestrians evacuate in response to an announcement towards an emergency exit located at the northern side (Figure 3).

The flood condition is generated by an inflow hydrograph formed by pairing a peak flow $Q_{peak}$ and an inundation duration $t_{inflow}$ as being major determining factor of flood risk. Four flooding conditions are investigated by fixing the volume of water that will be released into the shopping centre while proportionally doubling the discharge peak ($Q_{peak}$) and halving the duration of its occurrence ($t_{inflow}$). To ensure that the flood conditions generated are realistic, $Q_{peak}$ for 60 minutes of flooding is calculated according to the initial depth and velocity reported for the Norwich inundation case study reported by the UK Environment Agency (2006), which are $h_{inflow} = 1$ m (fixed) and $v_{inflow} = 0.2$ m/s. This corresponds to an initial pair ($Q_{peak}$, $t_{inflow}$) = (20 m³/s, 60 min), with $Q_{peak} = v_{inflow} h_{inflow} B$ and $B = 100$ m is the length of the inflow breach (Figure 3). The other three pairs representing increasingly more severe flood conditions are: ($Q_{peak}$, $t_{inflow}$) = (40 m³/s, 30 min), (80 m³/s, 15 min) and (160 m³/s, 7.5 min), respectively, as shown in Figure 4a.

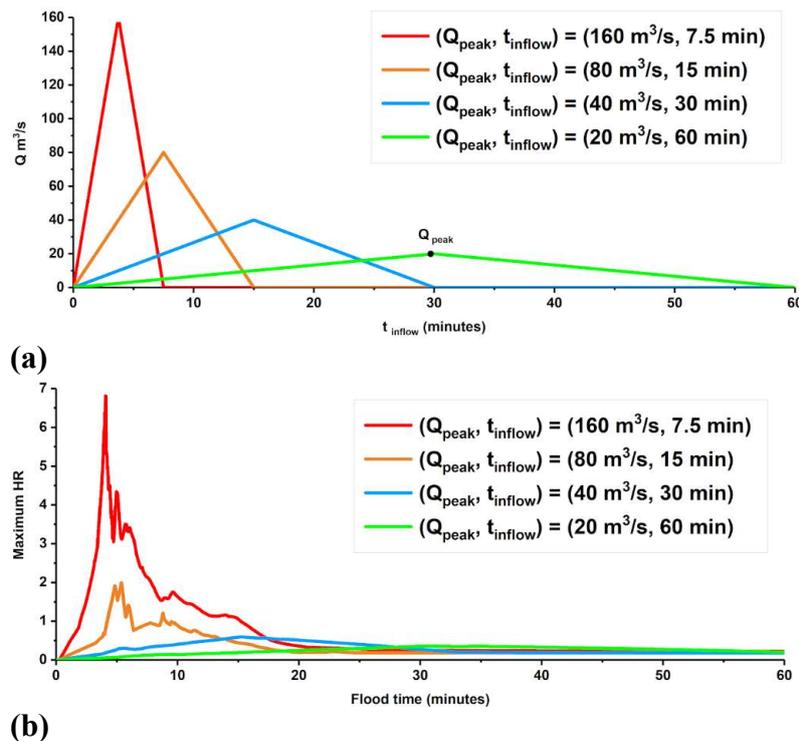

(a)

(b)

**Figure 4.** Flood conditions analysed for the shopping centre test: (a) hydrographs relative to four different ($Q_{peak}$, $t_{inflow}$) pairing; and (b) time history of the maximum HR as consequence of each inflow hydrograph.

To analyse flooding consequences relative to the four flood conditions, the hydrodynamic ABM is run for all four inflow hydrograph, with slip (numerical) boundary conditions for the northern side and wall (numerical) boundary conditions for the eastern and western sides (Figure 3), and $n_M$ = 0.01. Figure 4b shows the time history of the maximum HR for the four flood conditions during 60 minutes. With the pairs (20 m³/s, 60 min), (40 m³/s, 30 min) and (80 m³/s, 15 min), the maximum HR barely exceeds "1" over a very short duration. This indicates that these flood conditions can at worst lead to medium flood risk states. In contrast, the pair (160 m³/s, 7.5 min) leads to much higher range of maximum HR occurring over the first 10 minute of flooding. As being the worst-case scenario, only the flood condition based on the hydrograph (160 m³/s, 7.5 min) will be considered later on for the simulation of pedestrian moves and responses in/to floodwater.

The simulator is run for a grid of 128 × 128 flood agents interacting with a grid of 128 × 128 navigational agents where pedestrian agents can span and move (i.e. a fixed grid resolution of 2.59 m). For this study, a population of 1000 pedestrian agents walking at a resolution of around 0.5 m is selected, assuming a peak hour. When the domain is wet, the adaptive time-stepping mechanism of the hydrodynamic ABM is activated automatically and governs the simulation time-step under the CFL criterion (Wang *et al.* 2011); otherwise, a 1.0 s time-step is used by default. The pedestrian ABM is set to have a constant rate of 10 entering/leaving pedestrians per entrance/exit such that to maintain a constant population of the 1000 pedestrians before the flooding happens ($t$ = -5 min) while allowing pedestrians to spread all over the walkable area (Figure 3). As soon as flooding enters the domain, at $t$ = 0 min, the pedestrians become evacuees and the simulation is set to terminate when all evacuees leave the domain via the emergency exit (Figure 3).

Simulations are run by taking a diagnostic approach involving three *configuration modes* with systematic increase in the level of sophistication for the pedestrian behavioural rules:

- *'Mode 1'* only uses the simplified rules with constant walking speeds (Table 1);
- *'Mode 2'* integrates variable walking speeds using the empirical formula of Eq. (1);
- *'Mode 3'* further integrates the stability rules as described in Table 2.

In each run, the simulator is set to process and record every time step the information relevant to the water depth, velocity magnitude and HR values stored by the flood agents, and that of the pedestrian agents including their coordinates, HR-related flood risk states (Table 1), walking speed states (Eq. 1) and mobility states (Table 2).

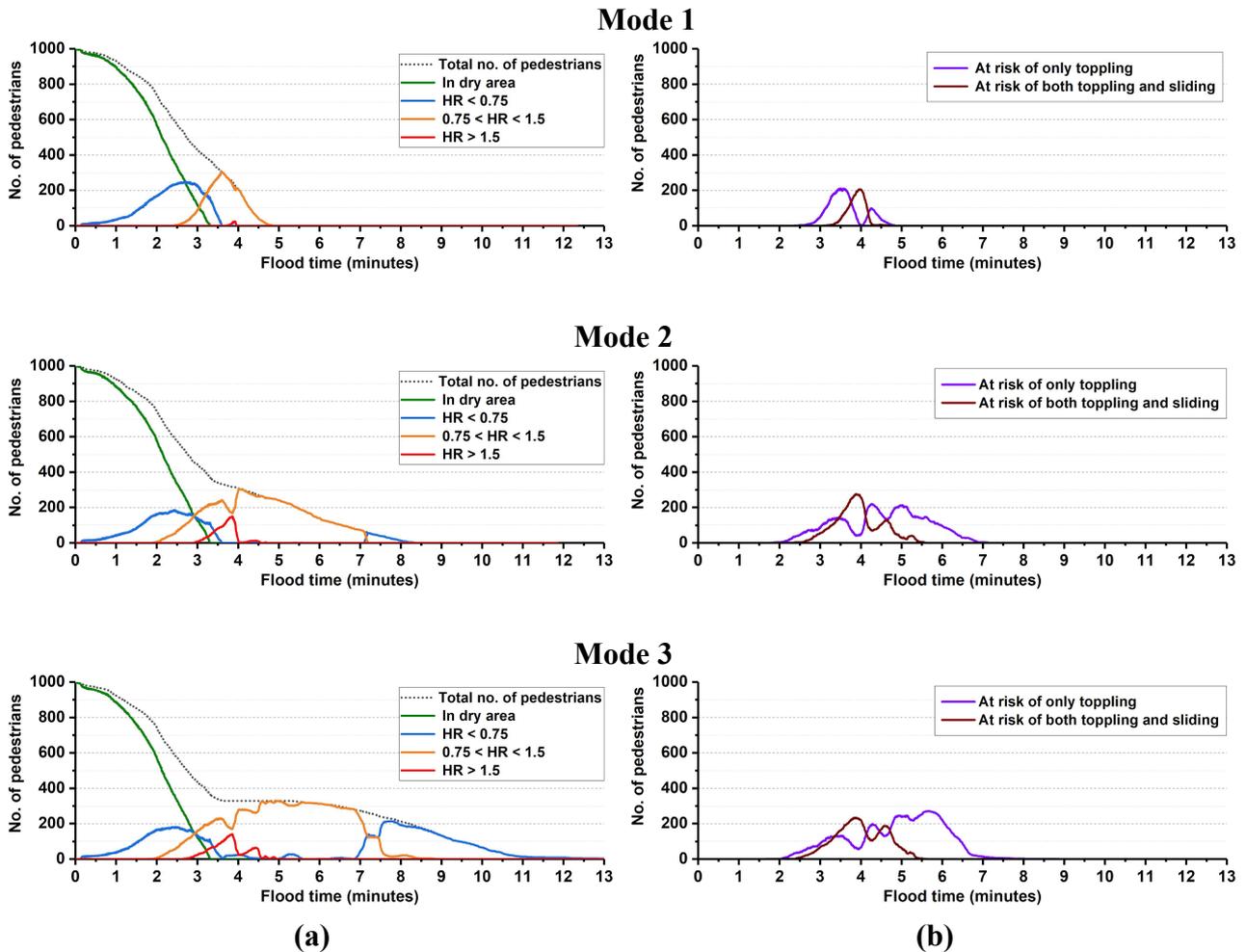

Figure 5. Number of evacuating pedestrians during the 13-minute flooding in simulation modes 1 to 3: (a) flood risk states in terms of local HR ranges (Table 1); and (b) states of unstable pedestrians under toppling and/or sliding (Table 2).

## 3. Numerical results

### 3.1 Analysis of flood risk and stability states

Figure 5 shows the time history of the number of evacuating pedestrians for the simulation modes 1 to 3 in terms of HR-related states (Figure 5a) and unstable states (Figure 5b). In mode 1 (Figure 5a, upper part), the simulator predicted an evacuation time of less than 5 minutes (Figure 5a, upper part, 'black' dotted line). In the first 2.5 minutes (Figure 5a, upper part, 'blue' line), the majority of pedestrians walk in dry areas (Figure 5a, upper part, 'green' line) while the rest walk in a low risk state of HR ≤ 0.75 (Figure 5a, upper part, 'blue' line) where they remain able to walk without losing stability (Figure 5b, upper part). After 2.5 minutes, 0.75 < HR ≤ 1.5 (Figure 5a, upper part, 'orange' line), indicating an increase in the flood risk state to medium to about 300 pedestrians. By 4 minutes, HR becomes higher than 1.5 (Figure 5a, upper part, 'red' line), imposing high to sever flood risk states on around 20 pedestrians. In the period of medium to severe flood risk, the number of unstable pedestrians increased, especially by 4 minutes where around 200 pedestrians are 'at risk of only toppling' or 'at risk of both toppling and sliding' (Figure 5b, upper part, 'dark red' and 'purple' lines).

In mode 2 (Figure 5, middle part), the simulator's predictions change significantly relative to mode 1. Now, it takes 3.5 minutes longer for the 1000 pedestrians to evacuate (Figure 5a, middle part, 'black' dotted line). Also, low risk state period spans a shorter duration of 2 minutes at the start of flooding but extends beyond 7 minutes, indicating that pedestrians can safely evacuate over last 1.5 minutes (Figure 5a, middle part, 'blue' line). Here, the medium to severe states are predicted to occur after 2 minutes and before 7 minutes, relatively affecting more walking pedestrians about 100 and 150, respectively. Mode 2 also leads to a prolonged period of unstable pedestrians, showing about 300 individuals 'at risk of only toppling' or 'at risk of both toppling and sliding' between 2 to 7 minutes of the flood time (Figure 5b, middle part).

In mode 3, even more prolonged periods of medium to severe states with unstable pedestrians are observed: the total evacuation time of 8.5 minutes, observed in mode 2, now increases to be 13 minutes (Figure 5b, lower part, 'black' dotted line). The duration of low flood risk states remains similar to as in mode 2 for the first 2 minutes, but span a longer duration after 7 minutes, thus indicating that the evacuating pedestrians would be in a low risk state after 8.5 minutes (Figure 5a, lower part, 'blue' line). During this period, before 2 minutes and after 8.5 minutes, pedestrians are identified stable that they are able to walk safely in floodwater (Figure 5b, lower part). Between 2 and 8.5 minutes of flood time, the flood risk seems to increase significantly, putting over 350 pedestrians in medium state and over 150 pedestrians in high to sever state even after 4 minutes of flood time. During this time window, the risk of having unstable pedestrians consistently rises, with more than 300 pedestrians 'at risk of only toppling' or 'at risk of both toppling and sliding'. It is useful to note that the risk of 'only sliding', although incorporated within the behavioural rules (Table 2), is not captured by the simulator. This could be associated with the flood hydrodynamics of this test case that does not entail enough shallow fast-flowing floodwater over the shopping centre.

Table 3. Relative changes in the outputs produced by the simulator under mode 1 and mode 2 relative to those produced under mode 3, quantified using the $R^2$ coefficient and $L^1$-norm error.

| Relative change relating to Mode 3 | | Dry | Low | Medium | High to severe | Toppling only | Toppling and sliding |
|---|---|---|---|---|---|---|---|
| Mode 1 | $R^2$ coefficient | 0.99 | 0.41 | 0.15 | 0.23 | 0.11 | 0.55 |
| | $L^1$-norm error | 1.91 | 44.04 | 81.44 | 13.52 | 47.69 | 26.66 |
| Mode 2 | $R^2$ coefficient | 0.99 | 0.49 | 0.78 | 0.90 | 0.84 | 0.95 |
| | $L^1$-norm error | 1.75 | 28.58 | 30.88 | 4.06 | 14.89 | 7.18 |

Moreover, the effective differences between the simulator predictions made with modes 1 and 2 relative to mode 3 are quantitatively assessed, using the R-squared ($R^2$) coefficient and $L^1$-norm error, which have the following expressions:

$$R^2 = \left[ \frac{\sum_{t=1}^{T}(D_t^{mode\,3} - \bar{D}^{mode\,3})(D_t^M - \bar{D}^M)}{\sqrt{\sum_{t=1}^{T}(D_t^{mode\,3} - \bar{D}^{mode\,3})^2 \sum_{t=1}^{T}(D_t^M - \bar{D}^M)^2}} \right]^2 \quad (4)$$

$$L^1\text{-norm error} = \frac{1}{N_s}\left(\sum_{t=1}^{T}|D_t^{mode\,3} - D_t^M|\right) \quad (5)$$

where $t$ denotes the current time and $T$ the output simulation time; $D_t^{mode\,3}$ is a data value at time $t$ obtained the simulator under mode 3, and $D_t^M$ refers to the data value at the same time $t$ obtained with either of the two other modes, $M \in \{Mode\,1, Mode\,2\}$. The values $\bar{D}^{mode\,3}$ and $\bar{D}^M$ represent time-averaged means, and $N_s$ is the size of a data time series. The $R^2$ coefficient takes values between 0 and 1, indicating stronger correlation with mode 3 results as it gets closer to 1. The $L^1$-norm error is more effective to quantify the average deviations relative to mode 3, and gets closer to 0 in line with reduced deviation. Table 3 lists the $L^1$-norm errors and $R^2$ coefficients quantifying the differences in terms of numbers of pedestrians predicted: in 'dry' areas, in 'low', 'medium', and 'high to severe' flood risk states, and with unstable states due to 'toppling only' and 'toppling and sliding'.

The $L^1$-norm error clearly indicates that the discrepancies among the simulator predictions with the different modes are significant, except for the number of pedestrians in dry areas. This is expected as the newly implemented behavioural rules are only relevant and activated for pedestrian agents in wet areas. As explored via Figure 5 a key reason leading to such large discrepancies is the major differences in evacuation times predicted under the three different modes. The $L^1$-norm errors also suggest that the predictions made by the simulator under mode 2 are closer to those made under mode 3, which is also expected as both modes 2 and 3 employed the same variable walking speed rules. This observation is clearer by analysing the range of the $R^2$ coefficients relative to mode 2, i.e. $0.49 \leq R^2 \leq 0.95$, suggesting that the evacuation patterns predicted using mode 2 are 49% to 95% similar to those under mode 3; whereas those under mode 1 yield results that are at very best 41% similar.

The results in Figure 5 also points out to potential different patterns for the spatial distribution of the pedestrians relative to the different simulation modes during their evacuation of the flooded shopping centre. To analyse the extent of difference in the predicted spatial distribution of pedestrians, the outputs relevant to the coordinate data of pedestrians are compared across the simulation mode 1 to 3. The analysis is performed after 4 minutes of flood time when the flood risk states were simulated high to severe in all the three simulation modes (Figure 5).

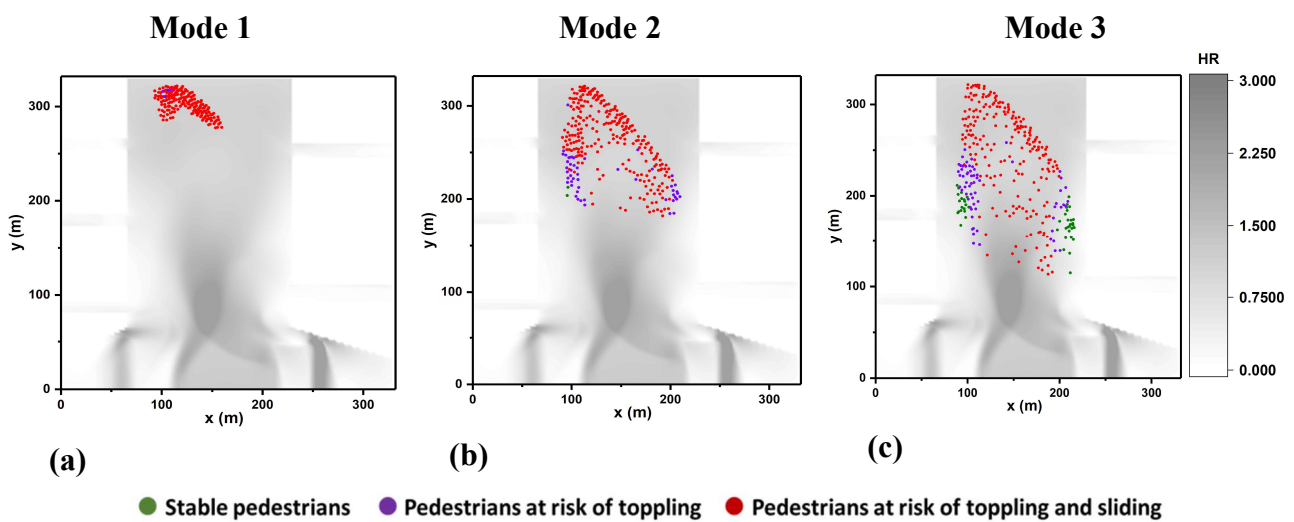

**Figure 6.** Location of the evacuating pedestrians (represented by points) at 4 minutes after flooding for the simulator runs with: (a) Mode 1, (b) Mode 2, and (c) Mode 3.

Figure 6 shows the location of the pedestrians, represented by dots, alongside 2D contour plots of the HR in the shopping centre representing flood extent for the three simulation modes. The different colours for the dots also indicate the stability states of the pedestrians. In mode 1 (Figure 6a), almost all the pedestrians are predicted to have unstable states of due to toppling only or toppling and sliding within a short distance of the emergency exit. In mode 2 (Figure 6b), similar pattern is observed but with much wider scattering from within the middle and at a larger distance from the emergency exit. This behaviour could be attributed to the variable walking speed, related to the flooding dynamics (Eq. 1), that the pedestrians could have with mode 2, as compared to the constant speeds used with mode 1 (Table 1). In mode 3 (Figure 6c), the evacuees are found to be even more spread away from the emergency exit relative to mode 2 as they become unstable (i.e. immobilised by toppling or toppling with sliding). Nonetheless, the simulator with mode 3 predicts a number of stable pedestrians evacuating around the sides of the open area where HR is relatively low.

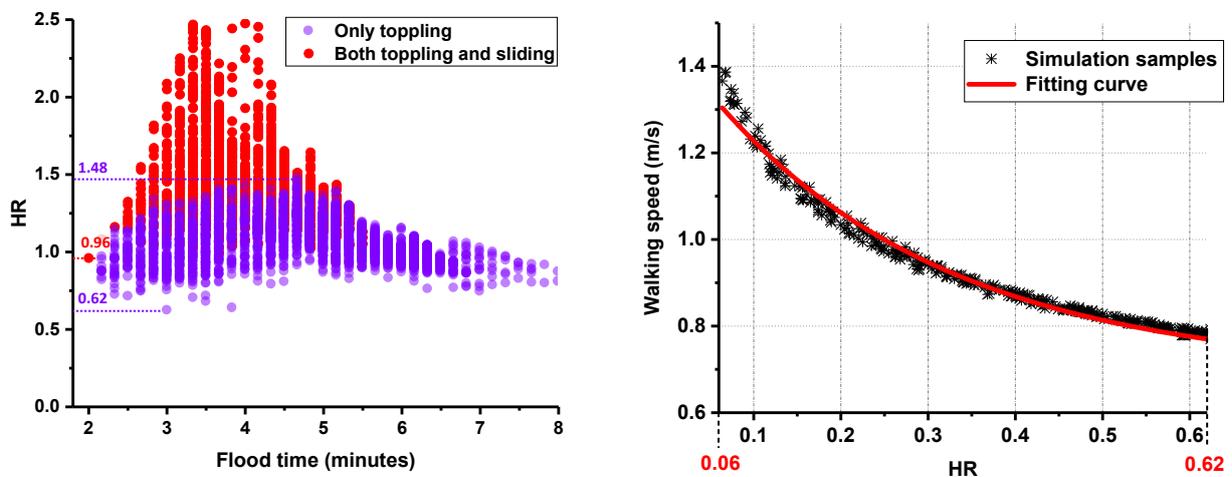

(a)  (b)

**Figure 7.** Relationship linking pedestrians' flood risk states in terms of HR ranges to their unstable states, and to their walking speed states when stable based on simulator's outcomes under mode 3: (a) distribution of unstable pedestrians' flood risk state of HR that are also identified at risk of 'toppling only' and of 'both toppling and sliding' after 2 and before 8.5 minutes of flood time; and, (b) distribution of stable pedestrians' walking speeds as function of HR identified before 2 and after 8.5 minutes.

### 3.2 Discussions and limitations

The results show that with increased level of sophistication in pedestrian behavioural rules, the simulator predicts increasingly prolonged evacuation time during, larger number of pedestrians with medium to severe flood risk states and of unstable pedestrians (Figure 5 and Table 3). Moreover, as the rules became more sophisticated, major differences in the spatial distribution of the pedestrians is identified, i.e. under mode 3 where their evacuation shows wider pattern and include larger number of stable pedestrians with low to medium flood risk states (Figure 6). These findings suggest that more realistic incorporation of human response dynamics into the flood risk analysis is likely to yield significantly different outcomes when planning evacuation times or aiming to pinpoint safest shelter areas and evacuation routes.

The spatial and temporal outcomes produced by the simulator have been further analysed to produce relationships linking flood risk states of pedestrians to their stability and walking speed states. Two sets of analysis have been performed under simulation modes 1 to 3, but only the results of simulation mode 3 are illustrated as they were found inclusive of all the possible outcomes identified with simulation mode 1 and 2. First analysis focused on charting the HR-related flood risk states of unstable pedestrians to identify how HR thresholds relate to 'only toppling' and 'both toppling and sliding' conditions. In the second set of analysis, the focus was given to chart the HR-related flood risk states, but on stable pedestrians, to identify how HR can be related to walking speed states.

The results of the first analysis are shown in Figure 7a, which illustrates the HR ranges of unstable pedestrians during 2 to 8.5 minutes of flood time: the 'purple' dots represent those unstable due to only toppling and the 'red' dots those unstable due to both toppling and sliding. The 'purple horizontal dotted lines' indicate the upper and lower limits of HR identified for toppling risk and the 'red' one indicates the lower limit of HR for the risk of both toppling and sliding. Pedestrians are found at risk of toppling only when $0.62 \leq HR \leq 1.48$, whereas they are at risk of both toppling and sliding when $HR \geq 0.96$. Figure 7b shows the plot of the pedestrian walking speed states versus HR, for $0.06 < HR < 0.62$, where the walking speed of pedestrians is identified to be affected by floodwater and the pedestrians remain stable (before 2 and after 8.5 minutes of flood time): As local HR increases from 0.06, the walking speed of pedestrians (initially 1.4 m/s when in a dry area) decreases down to 0.78 m/s at the threshold where they become instable (HR = 0.62). An explicit relationship linking HR to stable pedestrians' walking speed has been produced and is illustrated in Figure 7b. Based on an exponential fitting curve over the simulated results, this curve has yielded the best agreement with the simulation samples ($R^2 = 0.99$) and reads:

$$V_i = 0.69177 + 0.7762 \times 0.02466^{HR} \tag{6}$$

Overall, these sets of analysis on the simulation samples allows to identify: (i) thresholds for HR to directly estimate unstable states of pedestrians in floodwater; and, (ii) a formula to directly estimate walking speed states of stable pedestrians by only referring to HR. The identified thresholds and formula are expected to widen the utility of the HR metric (Environment Agency 2006; Kvočka *et al.* 2016; Willis *et al.* 2019; Costabile *et al.* 2020). For example, to gain more detailed insights on people mobility in floodwater for the same test case but different flood conditions or for other test cases that previously employed the HR metric. Moreover, the analysis reported here seems to imply that HR < 0.75 may be an overly optimistic recommendation as a low risk state ('safe for all', Environment Agency 2006) because the simulator still predicted a risk of toppling with HR > 0.62. Similarly, the lower limit of the medium risk state (i.e. HR > 0.75: 'dangerous to some', particularly children, Environment Agency 2006) may also be overly optimistic as the present simulator identifies a risk of toppling and sliding for adults for HR ≥ 0.96. These findings support conclusions made by other studies (Kvočka *et al.* 2016; Chanson & Brown 2018), pointing out to the need to quantify more conservative safety thresholds.

This study, however, just provide a starting point and has many limitations. One key limitation has been the lack of test cases with suitable observational data to validate against (e.g. video footages of an evacuation in flooded and populated area), irrespective of the fact that modelling behaviours and validation of ABMs is, by itself, a timely and grand challenge (An *et al.* 2020). Other limitations are related to the fact that the effects of pedestrian gatherings on local flood hydrodynamics were not incorporated, nor those due to having pedestrian of different age, gender, physical disability, *etc*.

## 4. Summary and conclusions

This work augmented the design of a flood-pedestrian simulator pertaining to increasing the level of sophistication of the behavioural rules governing the mobility of individual pedestrians in floodwater. The simulator couples validated hydrodynamic and pedestrian agent-based models (ABMs), allowing to concurrently outputs spatial and temporal outcomes on: the risk states of pedestrians in floodwater in relation to a commonly used flood Hazard Rating (HR) metric, and their states of walking speed when stable in floodwater or otherwise their unstable states due to toppling and/or sliding. Relative to the previous version of the simulator (Shirvani *et al.* 2020), pedestrians with various body height and weight are considered, as well as experimentally-valid behavioural rules to implement, across the ABMs, walking speed states or unstable states in floodwater. The augmented simulator was applied to reproduce a 'during flooding' evacuation scenario of a synthetic test case of a flooded shopping centre populated by a thousand pedestrians during a worst-case scenario flooding. The simulator runs were applied diagnostically under three configuration modes, while increasing the level of sophistication of the behaviour rules, to systematically analyse the relative changes in the outcomes across the modes. The analysis suggests much longer evacuation times as the behavioural rules became more sophisticated, and larger number of pedestrians with both higher flood risk and lower risk states due to the wider spread of pedestrians. Further processing to the simulator outcomes allow to usefully identify HR-related thresholds and formula to directly estimate the states of unstable pedestrians or otherwise their walking speed states. The HR safety thresholds, identified through simulations, are found to be slightly more conservative than those recommended in UK guidance documents (e.g. Environment Agency 2006), reinforcing alternative findings in published literature (e.g. Chanson & Brown 2018). Work is presently ongoing to explore the potential of the flood-pedestrian simulator to study and plan evacuation scenarios for a real site involving populated urban spaces.

**Acknowledgements and software accessibility**

This work was supported by the UK Engineering and Physical Sciences Research Council (EPSRC) grant EP/R007349/1. We thank Mozhgan Kabiri Chimeh and Peter Heywood from the Research Software Engineering (https://rse.shef.ac.uk/) group for providing technical support during the implementation of the flood-pedestrian simulator on FLAMEGPU. We also thank Charge Rougé, Simon Tait and the two anonymous reviewers for their careful reading and their very many insightful comments and suggestions, which greatly help the authors improve the quality of this paper.

The flood-pedestrian simulator software is available on DAFNI (www.dafni.ac.uk), where it can be run from a graphical interfaces and supported by a detailed 'run guide' document. To access the server on DAFNI before March 2021, an interested user needs to request a personal account by contacting info@dafni.ac.uk, or more specifically tom.gowland@stfc.ac.uk regarding the flood-pedestrian simulator. Further updates on ongoing developments related to the flood-pedestrian simulator can be found on www.seamlesswave.com/Flood_Human_ABM.